# Second-Harmonic Generation Tuning by Stretching Arrays of GaAs Nanowires

Grégoire Saerens[1*], Esther Bloch[1], Kristina. Frizyuk[2], Olga Sergaeva[2], Viola V. Vogler-Neuling[1], Elizaveta Semenova[3,4], Elizaveta Lebedkina[3], Mihail Petrov[2], Rachel Grange[1], Maria Timofeeva[1]

[*] Corresponding Author: gsaerens@phys.ethz.ch

[1] ETH Zurich, Optical Nanomaterial Group, Institute for Quantum Electronics, Department of Physics, 8093 Zürich, Switzerland

[2] ITMO University, Kronverkskiy prospect 49, 197101 St. Petersburg, Russia

[3] DTU Fotonik, Technical University of Denmark, 2800 Kongens Lyngby, Denmark

[4] NanoPhoton–Center for Nanophotonics, Technical University of Denmark, 2800 Kongens Lyngby, Denmark

**Abstract** - We present a wearable device with III-V nanowires in a flexible polymer, which is used for active mechanical tuning of the second-harmonic generation intensity. An array of vertical GaAs nanowires was grown with metalorganic vapour-phase epitaxy, then embedded in polydimethylsiloxane and detached from the rigid substrate with mechanical peel off. Experimental results show a tunability of the second-harmonic generation intensity by a factor of two for 30% stretching which matches the simulations including the distribution of sizes. We studied the impact of different parameters on the band dispersion and tunability of the second-harmonic generation, such as the pitch, the length, and the diameter. We predict at least three orders of magnitude active mechanical tuning of the nonlinear signal intensity for nanowire arrays. The flexibility of the array together with the resonant wavelength engineering make such structures perspective platforms for future bendable or stretchable nanophotonic devices as light sources or sensors.

Compound semiconductor nanowires (NWs) offer a unique optical platform for various photonic applications including light emission in both linear and nonlinear regimes.[1–3] The high refractive index of semiconductor NWs provides confinement and enhancement of electromagnetic fields without significant light absorption in the visible part of the spectrum and almost fully relieved of losses in the near infrared region.[4,5] Direct band gap III-V semiconductor nanostructures such as GaAs are known as efficient sources of luminescent signal[3] and possess a strong optical nonlinearity allowing for enhanced nonlinear optical processes including second-harmonic generation (SHG).[6–8] The enhancement of the SHG signal has already been extensively studied in the past years. In the recent work of D. de Ceglia et al. a SHG conversion efficiency up to $10^{-5}$ can be expected for a single GaAs NW by resonantly coupling the pump and the second-harmonic (SH) fields.[9] S. Liu et al. demonstrated how overlapping magnetic dipoles in a NW array leads to an enhancement of the SHG conversion efficiency by four orders of magnitude in comparison to an unpatterned GaAs thin film.[10] While passive systems based on semiconductor nanoantennas and metasurfaces have already shown very effective light conversion and emission, the dynamical control and tunability of their optical properties is highly desirable for real-time modulation of optical signals. One of prospective directions are mechanically tunable systems for which their optical properties can be reconfigured by applying a mechanical force.[11,12] GaAs NWs are good candidates for various optomechanical systems as being obtained with a bottom-up or a top-down fabrication approach, they still can be released using wet etching,[13] peeling off,[14] transfer printing[15] or mechanical exfoliation[16] methods. By integrating them with various flexible media, several applications have been demonstrated as bendable or stretchable light sources,[13,16] sensors,[17,18] lenses,[19] electronic components,[20,21] optical filters[22,23] or photodetectors.[23,24]

While flexible optical structures have been utilized for tuning the linear optical properties of dielectric systems,[13,16–24] the mechanical control over the nonlinear signal generation has not been demonstrated yet. In this work, we show how to overcome the fixed geometry constraint for efficient light conversion with mechanical active tuning of the SHG intensity in NW arrays embedded into a flexible polymer film. The NWs were grown on a silicon substrate by metalorganic vapour-phase epitaxy and detached from the substrate with a sharp blade. We achieved up to 30% uniaxial tensile stretching and a mechanical tunability of the SHG by a factor of almost two. The SHG conversion efficiency at 0% stretching was around $10^{-9}$ $W^{-1}$. These experimental results were further supported with simulations of SHG conversion efficiency tunability of an uniaxially stretched array of NWs with different lengths and diameters and for an incoming polarization parallel or perpendicular to the stretching direction. We provide guidelines with range of design parameters that can be targeted to obtain a maximised tunability. The calculations predict maximum values of three orders of magnitude with theoretical SHG conversion efficiency up to $10^{-3}$ $W^{-1}$. We believe these hybrid features of a NW array within a stretchable matrix are convenient for further applications as light sources or light sensors in biology or as key components for flexible and wearable display technologies. [25–27]

First, we provide the modelling used for the linear and nonlinear optical response of a 2D NW array. Calculations were performed with a finite element model for an ideal periodic structure. The rectangular unit cell is composed of polydimethylsiloxane (PDMS) and of a single GaAs NW with specific length L and diameter D in the centre (Figure 1a). At rest, the side lengths of the square unit cell are equal (Figure 1b). Under uniaxial stretching one side length of the rectangular unit cell is varied (x-direction), while the other is fixed (y-direction), see also Figure 1c. The pitch p is defined as the distance between neighbouring NWs in x-direction (see Methods for more details).

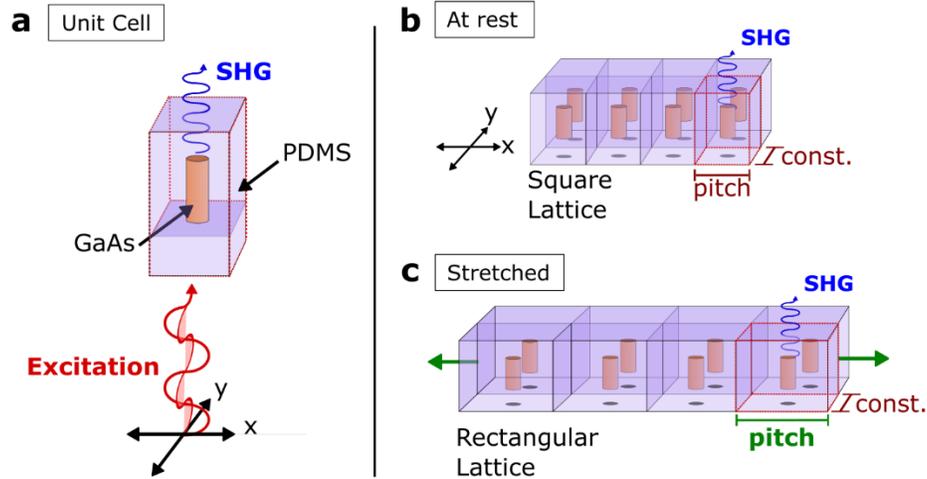

Figure 1. (a) Schematic of the periodic structure that is implemented in the finite element model. It consists of a unit cell with PDMS and a NW of specific length and diameter in the centre. (b) NW array at rest configuration and (c) under uniaxial stretching, for which the size of the unit cell varies only in one direction. A plane wave with linear polarization (x- or y-polarization) excites the GaAs NWs, generating SH.

The SHG conversion efficiency $\sigma$ is calculated with 1/W units as the generated SH peak power divided by the excitation peak power squared $\sigma = \frac{P(2\omega)}{P^2(\omega)}$, similarly as in the work of M. Celebrano et al.[28] The optical properties are governed by the lattice modes of the NWs array formed by the Mie resonances of individual cylinders. These modes impact greatly the SHG conversion efficiency but depend on the NWs length and diameter, on the pitch and on the excitation polarization. The band structure of the modes and the corresponding electric field distributions for a square lattice with a pitch of 600 nm (see Figure 2a) show four branches with preferable x- and y-polarization directions and a degeneracy at the centre of the Brillouin zone (Γ-point). The modes 1 and 2 play the most important role as their Q-factor value obtained in numerical modelling is around Q~300. Modes 3 and 4 are low Q-factor modes with Q~5 and so do not contribute into the overall field enhancement (see Figure S1 in the Supporting Information). This can also be seen in the field enhancement spectra (see Figure 2b), where a single peak is observed exactly at the wavelength corresponding to the Γ-point of modes 1 and 2.

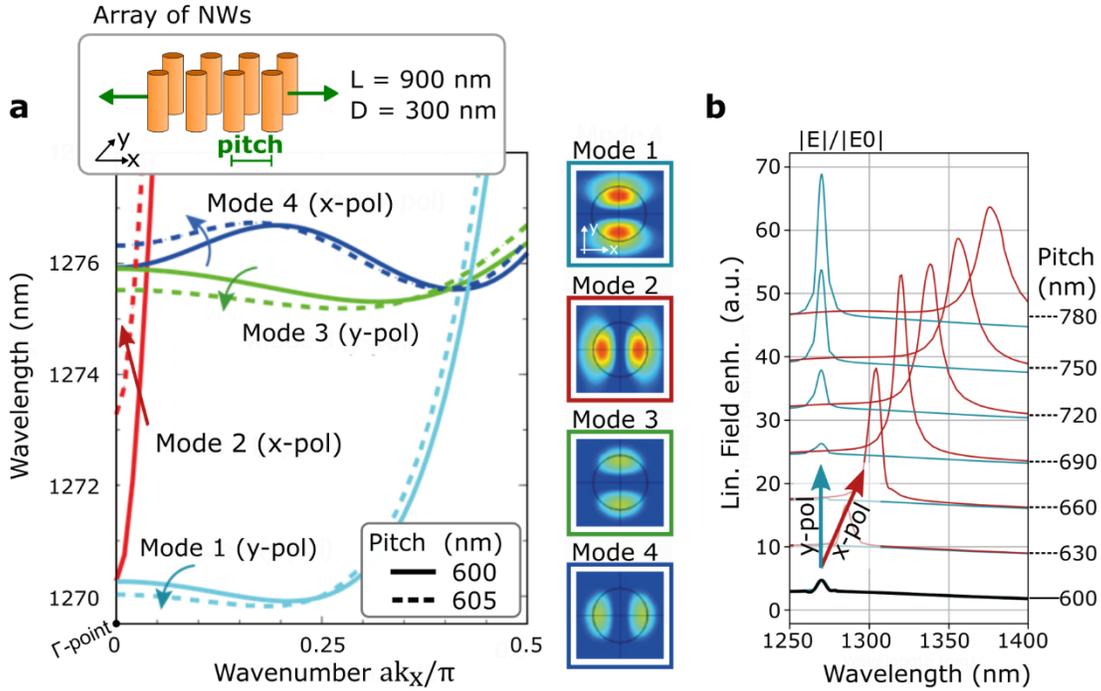

Figure 2. (a) Photonic band structure in an array of NWs with length L = 900 nm and diameter D = 300 nm. Dispersion of modes contributing to the transmission and reflection spectra are shown for a square grating with pitch p = 600 nm (solid line) and for a rectangular grating with pitch p = 605 nm (dotted line). The inset shows the electric field distribution demonstrating x- and y-polarization of the modes. (b) Linear field enhancement, expressed as the ratio of electric field intensity inside the structure to the incoming field power. The resonance red-shifts only for x-polarization similarly as predicted from the band structure.

By stretching along the x-direction, which results in a pitch increase from 600 nm to 605 nm, the degeneracy at the Γ-point is split and the frequency of the modes are shifted. However, Mode 2 (x-pol) has a significantly steeper dependence on the k-vector than Mode 1 (y-pol) resulting in a different sensitivity to perturbations along the x-direction and, thus, a different spectral shift: only 0.3 nm for Mode 1, while for Mode 2 the shift is on the order of 3 nm (see Figure 2a). This is in perfect agreement with the simulations of the field enhancement spectrum, expressed as the ratio of the electric field intensity inside the structure to the incoming field power (see Figure 2b). We observe a strong spectral shift for an excitation with x-polarization (corresponding to Mode 2) and a weak spectral shift for an excitation with y-polarization (corresponding to Mode 1). Further analysis of modes is presented for a NW array in section S1 (see Figures S2-S4) and for a single NW in section S2 (see Figures S5, S6) in the Supporting Information.

The SHG conversion efficiency spectra are shown in Figure 3a,b for an array of NWs with length L = 900 nm, diameter D = 300 nm and pitch p from 600 nm to 780 nm. These values are similar to the sample of NWs array that will be presented later in the experimental section. At rest configuration, the spectrum shows a dominant lattice resonance around λ = 1270 nm (blue curve in Figure 3a,b). This does not depend on the incoming polarization excitation being in the x or y direction as the periodic structure is symmetric under a 90° rotation along the light propagation axis. The variation of the pitch in one direction with x-polarization excitation leads first to an increase and then a decrease in the amplitude of the lattice resonance (Figure 3a). Additionally, the position shifts for the x-polarization due to the overlap of the individual Mie resonances of the single NW elements (see Figure S2 in the Supporting Information). An SHG conversion efficiency tunability can be calculated as $G = \frac{\sigma_s}{\sigma_0}$, where $\sigma_s$ and $\sigma_0$ are the SHG conversion efficiencies measured at the same wavelength but at different pitches, between p = 630 and p = 780 nm for the first (stretched) and at pitch p = 600 nm for the latter (at rest configuration). Around the lattice resonance a maximum value of G ≈ 225 for the excitation with x-polarization is found, while for the excitation with y- polarization this value is G ≈ 17. The tunability factor $G$ was calculated (given with a logarithmic colour scale) for arrays with NWs of different lengths L and diameters

D, as shown in Figure 3c,d. For NWs with length around L = 1000 nm and diameter around D = 300 nm, the value of G is three orders of magnitude for the x-polarization (see Figure 3c). Similarly, for NWs with lengths L = 900 nm and diameter D = 260 nm, the value of G is two orders of magnitude for the y-polarization (see Figure 3d).

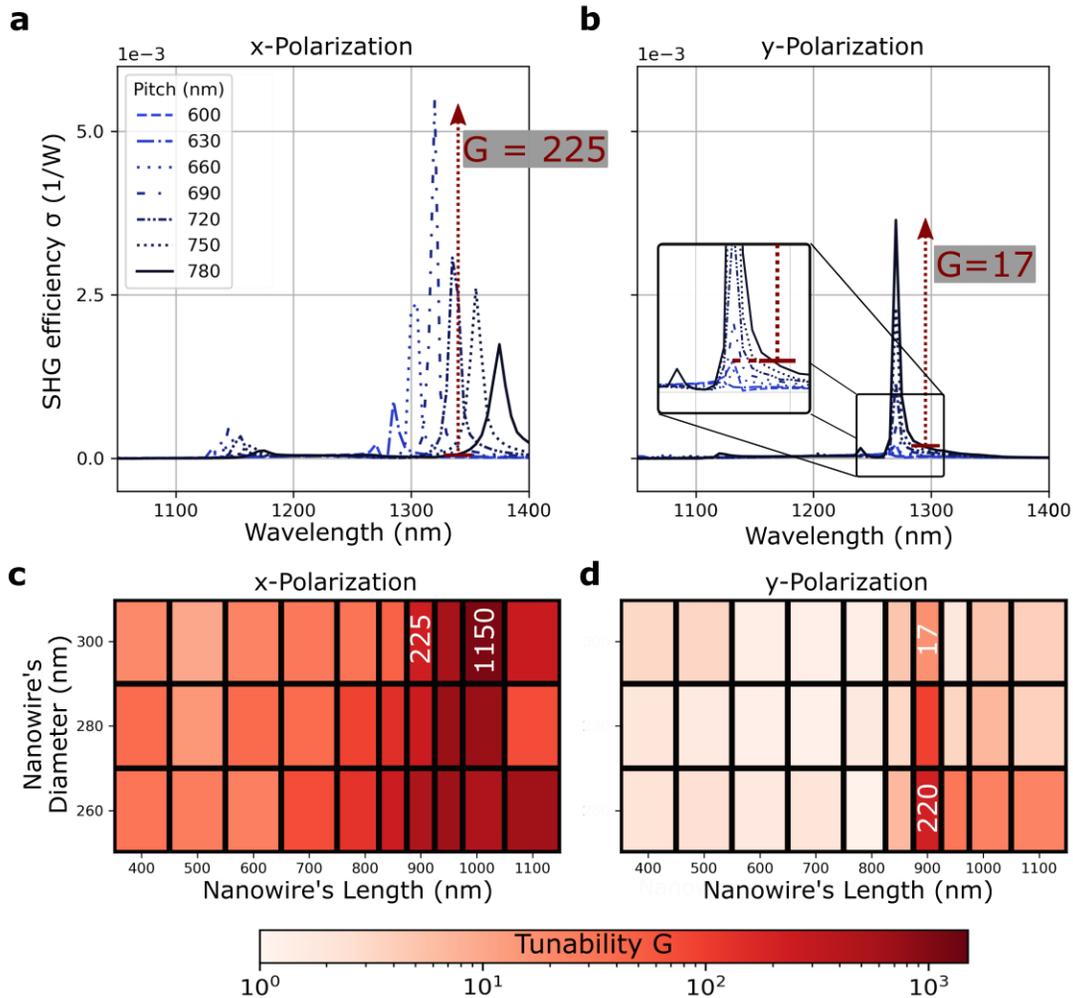

Figure 3. Simulated SHG conversion efficiency spectrum for an array of NWs with length L = 900 nm and diameter D = 300 nm is given for different pitch values from p = 600 nm (at rest configuration) to p = 780 nm (30% stretched). These values correspond also to the samples array of the experiments. Nonlinear efficiency calculated for an excitation (a) with x-polarization, (b) with y- polarization. The maximum nonlinear signal tunability for x-polarization is G = 225, and for the y-polarization is G = 17. (c,d) Calculated G values for different NWs lengths and diameters,

plotted in a logarithmic scale. For an excitation with x-polarization, G is at least three orders of magnitude for slightly longer NWs, while with y-polarization, G is at least two orders of magnitude for slightly thinner NWs.

The higher values for an excitation with x-polarization can be explained by the red-shift of the lattice resonance as the SHG intensity at rest is almost zero at the resonance position of the stretched case. We believe even higher values can be reached by optimizing the pitch and varying with finer steps the NWs length and diameter. NWs shorter than 800 nm do not show high G values because the dominant lattice resonance is broader than for longer NWs.

We present experimental results that support the above simulations. The different fabrication steps of a stretchable periodic structure are schematized in Figure 4a. A dielectric mask was used on silicon to grow arrays of NWs with metalorganic vapour-phase epitaxy (MOVPE). Scanning electron microscope images of the NWs were taken before a layer of flexible PDMS material was deposited (see Figure 4b). Embedded NWs were finally mechanically extracted using a razor blade. More details about the fabrication procedure are given in Methods (see also Section S3 in the Supporting Information). The flexible structure of NWs allows mechanical tensile stretching up to 30%, which is limited by the size of the sample compared to the setup. The SHG conversion efficiency was measured in transmission with the excitation wavelength in the region from $\lambda = 1050$ nm to $\lambda = 1350$ nm. The setup is shown in Figure 4c and described in Methods.

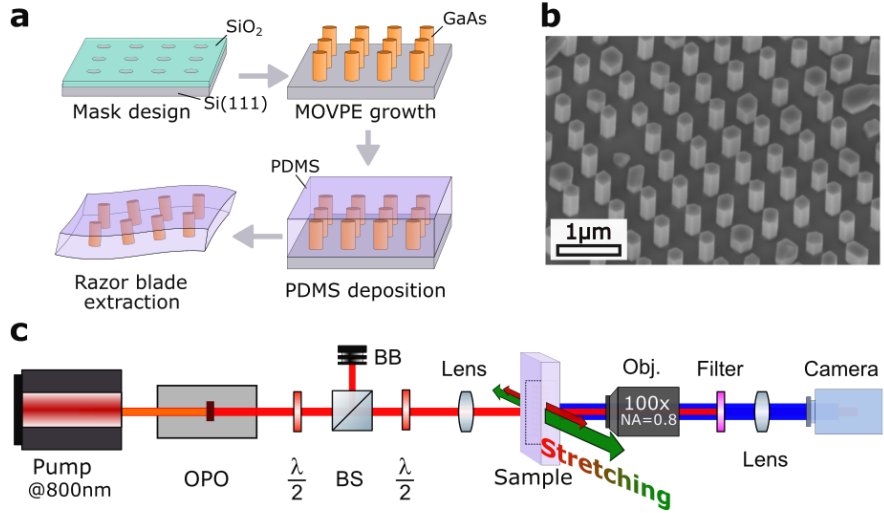

Figure 4. (a) Schematized fabrication steps of the stretchable sample. A SiO$_2$ mask was fabricated, NWs were grown selectively on silicon with MOVPE, a layer of flexible PDMS material was deposited and embedded NWs were finally mechanically extracted using a razor blade. (b) Scanning electron microscope image of a GaAs NWs array selectively grown on Si. We measured a pitch of p = 600 nm. (c) Schematic transmission setup to measure the SHG conversion efficiency. OPO: optical parametric oscillator, λ/2: half-wave plate, BS: beam splitter, BB: beam block. The first λ/2 and the BS are used to select the power, while the second λ/2 is used to rotate the polarization.

Two regions containing short NWs (around L = 500 nm) and long ones (around L = 900 nm) were selected for linear and nonlinear optical characterization (see Figure S8 in the Supporting Information). The experimental results for the transmission show good agreement with the simulation results, as shown in section S4 of the Supporting Information. The SH signal was measured at 10 different spots in each of the two regions (see Figure 5a,b) and for an excitation with x-polarization (solid line) and y-polarization (dotted line). The characterization of the SHG signal is shown in section S5 of the Supporting Information. The spectra at rest (blue) are similar for the two polarizations, which is expected as the array is symmetric with respect to a 90° rotation along the light propagation axis. The resonance is broader than in the above simulations due to the not completely optimized physical bottom-up growing process that led to a non-negligible distribution of NWs lengths and diameters, which we could evaluate from SEM images. By measuring this local distribution and extrapolating numerically the SHG efficiency from arrays of NWs with different lengths and diameters, we calculated a single spectrum of the SHG conversion efficiency (see Methods and Figures S8,9 in the Supporting Information). Figure 5c shows this calculated spectrum for a region with shorter NWs, respectively Figure 5d shows for a region with longer NWs. The simulations indicate a bigger tunability of the conversion efficiency for longer NWs than for shorter ones, as well as a red-shift of the resonance to $\lambda = 1260$ nm for the x-polarization. These two specific aspects are also visible in the experimental results shown in Figure 5b.

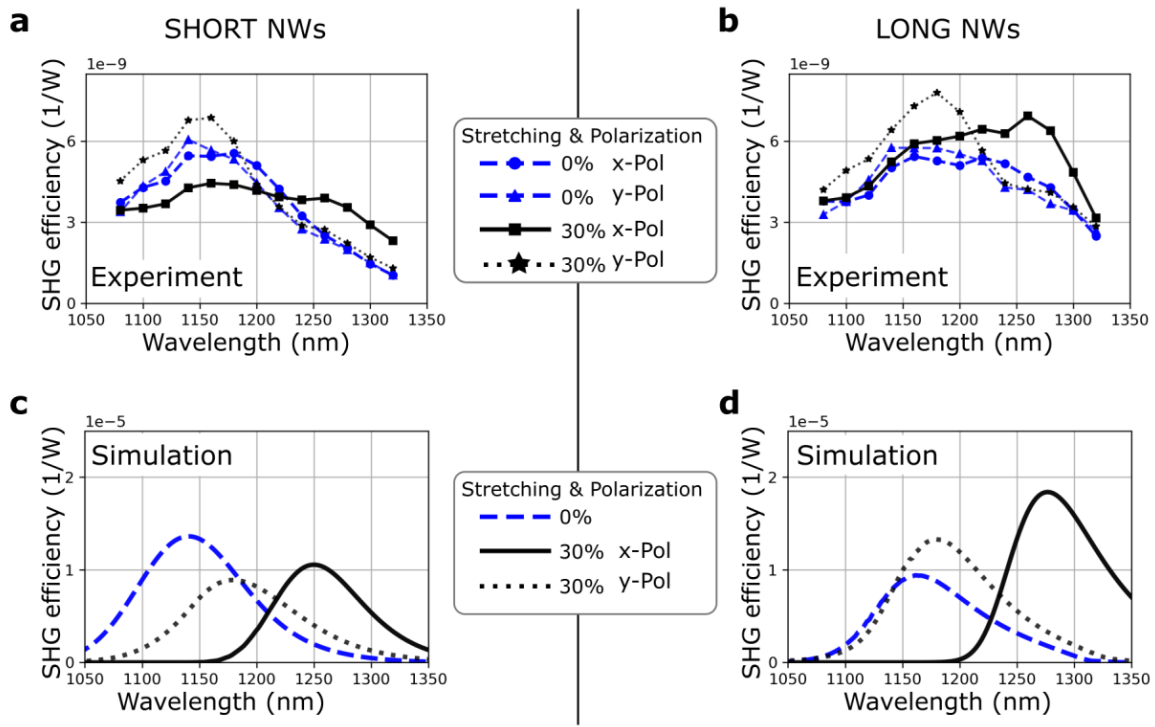

Figure 5. Experimental results of measured SHG conversion efficiency spectrum for the two regions of (a) short (around L = 500 nm) and (b) long (around L = 900 nm) NWs. In blue, measured when at rest and in black when stretched by 30%. The SHG conversion efficiency spectrum was measured at 10 different spots for the x-polarization (solid line) and y-polarization (dotted line). (c,d) Simulation results calculated taking into account the specific

distribution of NWs size. For longer NWs, an increase in the SHG efficiency as well as peak position shift for an excitation with x-polarization was visible in the experimental and simulations results.

The measured tunability of the SHG conversion efficiency is not as strong as calculated with our simulation model. This can be explained by the sample imperfections as the NW arrays possess a significant distribution of sizes and also 10% of not-fully grown NWs (see Figure 4b and Figure S8 in the Supporting Information). In the simulations, the infinitely periodic structure was excited uniformly and the SHG was collected with all angles in the forward direction. However, in the experiment, the beam had a radius around 5 μm, exciting around 100-200 NWs, and the objective only collected in forward direction with a numerical aperture of 0.8. Another important aspect to be mentioned is that the interaction of NWs with different sizes was not considered in the simulations, while in the sample, NWs with slightly different lengths were located next to each other. We measure the SHG conversion efficiency like this $\sigma = \frac{P(2\omega)}{P^2(\omega)} = 10^{-9}$ W$^{-1}$, or equivalently as a unitless efficiency $\frac{P(2\omega)}{P(\omega)} = 10^{-6}$ for an average excitation power of 5 mW/cm$^2$. The SH power $P(2\omega)$ measured after the sample and the excitation power at the sample $P(\omega)$ take into account the transmission losses from the optics. This experimental SHG conversion efficiency value is similar to single nanostructures or periodic arrangements. [10,29–31]

In conclusion, we demonstrate how arrays of GaAs NWs can be efficiently used to mechanically tune the SHG intensity in the near infrared. We fabricated a periodic structure based on patterned GaAs NWs embedded in PDMS. Our numerical calculations show that a tunability of the SHG conversion efficiency of at least 3 orders of magnitude is possible for an excitation polarization along the uniaxial stretching direction. Experimental results show an SHG tunability of almost 2 for 30% mechanical stretching and our developed numerical model indicate that the NWs size distribution is reducing the nonlinear optical performances. We calculate the SHG conversion efficiencies and tunabilities for an array with NWs of specific lengths and radiuses and believe these values can be used as a guideline for stretching capabilities of GaAs NWs arrays. Experimental improvements to obtain stronger resonances are possible by optimizing the epitaxial growth, or using a top-down approach. The SHG tunability can further be increased by applying a bigger mechanical stretching, especially for an excitation with y-polarization, or by stacking different layers of flexible NWs array on top of each other.

Methods

**Nanowire Growth**

A 60 nm thick $SiO_2$ mask was fabricated. GaAs nanowires were grown on pre-patterned Si (111)-oriented wafer in a low-pressure (80 mbar) MOVPE Turbodisc® reactor with hydrogen as a carrier gas. The growth temperature was 750° C. Trimethylgallium with flow rate of 7.8e-6 mol/min and arsine with flow rate of 2.23e-3 mol/min were used as group III and group V precursors, respectively. The pitch between the NWs was around 600 nm, their radius was around 130 nm, while their length varied from 500 nm (NWs in the centre) to 1300 nm (NWs in the edge) with a local distribution in the length of ± 100 nm.

**Sample Fabrication**

Polydimethylsiloxane (PDMS) was fabricated from mixing a base and an agent with mass ratio 10:1 (SYLGARD® 184, silicone elastomer kit, Dowsil). We added Toluene (ACS reagent >99.7% (GC), Fluka) to change the density of the polymer (20% of total mass) and to achieve better stretching properties. After pouring the PDMS, we waited 10 min to allow the PDMS to penetrate between the NWs. The PDMS was spin coated (1000 rpm, 1000 rpm/s, 3 min) to form a thin layer embedding the grown NWs. It was heated to 80 °C for 1 hour and cooled for 3 hours. After that the thin layer with NWs inside was mechanically detached from the Si substrate by using a razor blade. An additional layer of PDMS was dropped on top and spin coated (100 rpm, 1000 rpm/s, 10 min) to form a large flexible sample. It was heated to 80 °C for 1 hour, cooled for another 3 hours and a square (25 cm x 10 cm) was cut out with a blade.

**Optical Measurements**

The SHG intensity spectrum was measured with a fully-automated homemade nonlinear microscope system as illustrated in Figure 4c. A pulsed light with tunable wavelength was generated by a Ti:Sapphire Laser system (Chameleon Ultra II, Coherent) combined with an optical parametric oscillator (Compact OPO, Coherent). The pulses wavelength was tuned from 1050 nm to 1350 nm with a duration typically around 200 fs, a repetition rate of 80 MHz and an average power at the sample of 5 mW. The beam was focused to a beam radius of 5 µm on the sample with a lens (A240TM with NA = 0.5, Thorlabs) and collected with a 100x objective (LMPlanFL N with NA = 0.8, Olympus). The sample was firmly attached on both sides and stretching was achieved by translating the two sample holders in opposite directions (precision of translation is 0.1mm for a 15 mm rest length). The signal was focused with a convex lens (la1461 with f = 250 mm, Thorlabs) onto a sCMOS camera (Zyla 4.2, Andor) and the SH signal was separated using two high-pass filters.

**Simulation model**

Calculation were performed with a finite element method (COMSOL Multiphysics). An infinite array was defined through its unit cell which consisted of a rectangle out of PDMS material (refractive index around $n(\omega) \approx 1.4$), with perfectly matched layers on top and bottom, and a GaAs NW in the middle (refractive index around $n(\omega) \approx 3.4$). The bottom side of the unit cell was set as an entry port for a plane wave excitation while the top side was a collection port for the SHG intensity. The electric field for the linear scattering regime was calculated for an incoming linearly polarized plane wave with an intensity of 30 GW/cm2, which corresponded to the peak power at the sample in the experimental setup. The induced nonlinear polarization was given by the $\chi^{(2)}$ tensor for Zinc-Blende GaAs, for which the biggest component $\chi_{36}$ = 370 pm/V. Since the NWs were grown along the (111) direction, the tensor was also rotated following tensor rotation calculations.[32]

**Accounting for NWs size distribution**

We developed another approach for which we fitted the main resonance from the simulation spectrums with a Gaussian peak and extrapolated the amplitude, the position and the width for arrays of NWs with different lengths and diameters. It was then possible to come up with a single and smoother approximation spectrum by adding the extrapolated Gaussian peaks with their extrapolated parameters as well as with the corresponding weight given from the (continuous) size distribution (see also Figure S9 in the Supporting Information).

**Author Contributions**

The manuscript was written through contributions of all authors. M.T., R.G. and M.P. designed the experiment. M.T. and G.S. built the setup and conducted the nonlinear experiments. M.P., O.S, K.F., E.B. and G.S. performed the different numerical simulations. E.L. and E.S. fabricated the arrays of GaAs NWs. G.S. prepared the flexible sample with the help of M.T. and V.V.-N. M.T., R.G., M.P. and G.S analysed the data. V.V.-N. and G.S. performed the transmission measurements. M.T., R.G., M.P and G.S. wrote the manuscript. All authors have given approval to the final version of the manuscript.

**Acknowledgments**

We thank E. Denervaud for taking SEM images of the structure as well as M. Tchernycheva and N. Amador for showing us the peeling off technique applied for the fabrication of a flexible structure. The authors thank the Scientific Centre for Optical and Electron Microscopy (ScopeM) of Eidgenössische Technische Hochschule (ETH) Zürich. This work was supported by the Ambizione Grant No. 179966, the Danish National Research Foundation Grant No. DNRF147 - NanoPhoton. This work was supported by the Swiss National Science Foundation Grant 179099 and 150609, the European Union's Horizon 2020 research and innovation program from the European Research Council under the Grant Agreement No. 714837 (Chi2-nano-oxides).

**Conflict of interest**

The authors declare no conflict of interest.

**Abbreviations**

SHG, second-harmonic generation; SH, second-harmonic; NW, nanowire; PDMS, Polydimethylsiloxane; MOVPE, metalorganic vapour-phase epitaxy; SEM, scanning electron microscopy.

# Supporting Information

# Second-Harmonic Generation Tuning by Stretching Arrays of GaAs Nanowires

Grégoire Saerens[1,*], Esther Bloch[1], Kristina. Frizyuk[2], Olga Sergaeva[2], Viola V. Vogler-Neuling[1], Elizaveta Semenova[3,4], Elizaveta Lebedkina[3], Mihail Petrov[2], Rachel Grange[1], Maria Timofeeva[1]

[*] Corresponding Author: gsaerens@phys.ethz.ch

[1] ETH Zurich, Optical Nanomaterial Group, Institute for Quantum Electronics, Department of Physics, 8093 Zürich, Switzerland

[2] ITMO University, Kronverkskiy prospect 49, 197101 St. Petersburg, Russia

[3] DTU Fotonik, Technical University of Denmark, 2800 Kongens Lyngby, Denmark

[4] NanoPhoton–Center for Nanophotonics, Technical University of Denmark, 2800 Kongens Lyngby, Denmark

## Contents

## S1. Details on field distributions and resonances of Nanowire arrays

We first show in Figure S1 the Q-factor resonance corresponding to Figure 2 in the main text. Mode 1 & 2 possess much stronger Q-factors of around 300 than Mode 3&4. Mode 3 and 4 are low Q-factor modes with Q ~ 5 and so do not contribute into the overall field enhancement.

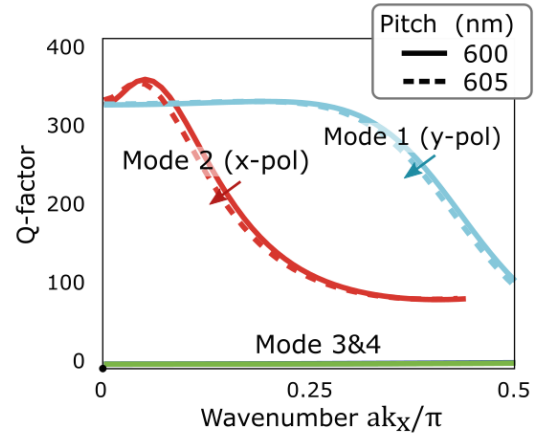

Figure S1. Q-factor corresponding to the band diagram shown in Figure 2 in the main text. Mode 1 & 2 possess much stronger Q-factors of around 300 than Mode 3 & 4 with Q ~ 5.

We also study the optical resonance supported in an array of NWs with length L = 900 nm, diameter D = 300 nm for which the second-harmonic generation (SHG) tunability under stretching is strongest. The linear transmission spectrum and the SHG conversion efficiency spectrum for this NWs array are given in Figure S2a-b. The resonance at $\lambda_{res}$ = 1270 nm wavelength indicated by the arrow is visible in the transmission and SHG spectrum. In Figure S2c-d we plotted the electric and magnetic field intensity distributions $|E(2\omega)|^2$ and $|H(2\omega)|^2$ (normalized to the incoming electric field) at a resonant wavelength of $\lambda_{res}$ = 1270 nm and a non-resonant wavelength of $\lambda_{nr}$ = 1240 nm. The second-harmonic (SH) electric field is the strongest at the edges of the NW while the SH magnetic field is the strongest inside the NW. Inside the NW at resonant wavelength, we can observe five nodes in the SH electric field distribution and six nodes in the SH magnetic field distribution, which may indicate a multipole or Fabry-Perot resonance.

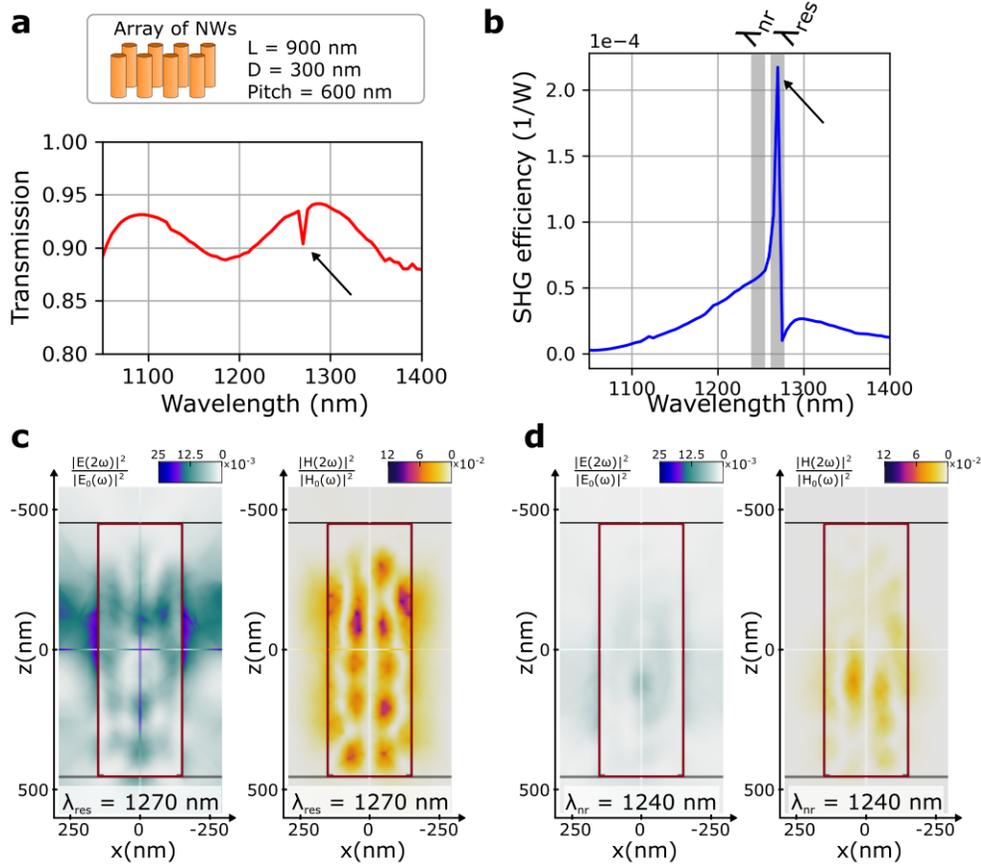

Figure S2. (a) NW Array with transmission spectrum and (b) SHG conversion efficiency. A resonance for high SHG generation is found at wavelength 1270 nm. (c,d) Relative intensity enhancement of the generated SH electric and magnetic fields at resonant wavelength $\lambda_{res}$ = 1270 nm and out of resonance at $\lambda_{nr}$ = 1240 nm. Five nodes in the SH electric field distribution and six nodes in the SH magnetic field distribution are visible inside the NW.

In Figure S3, we compare the linear electric field distributions (normalized to the incoming electric field) at rest configuration, when the pitch p = 600 nm, (see Figure S3a,b) with the fields at high resonance when the pitch p = 690 nm for x-polarization (see Figure S3c,d) and p = 780 nm for y-polarization (see Figure S3e,f). The field distributions are shown at the xy-plane with the NW in the centre. We observe in the stretched case the same linear electric field distributions (Figure S3b,d,f) as shown in Figure 2a with mode 1 (y-polarization) and mode 2 (x-polarization). The electric field corresponding to mode 2 at resonance (see Figure S3d) is also distributed in y-direction to the neighbour NW, indicating a strong coupling between NWs.

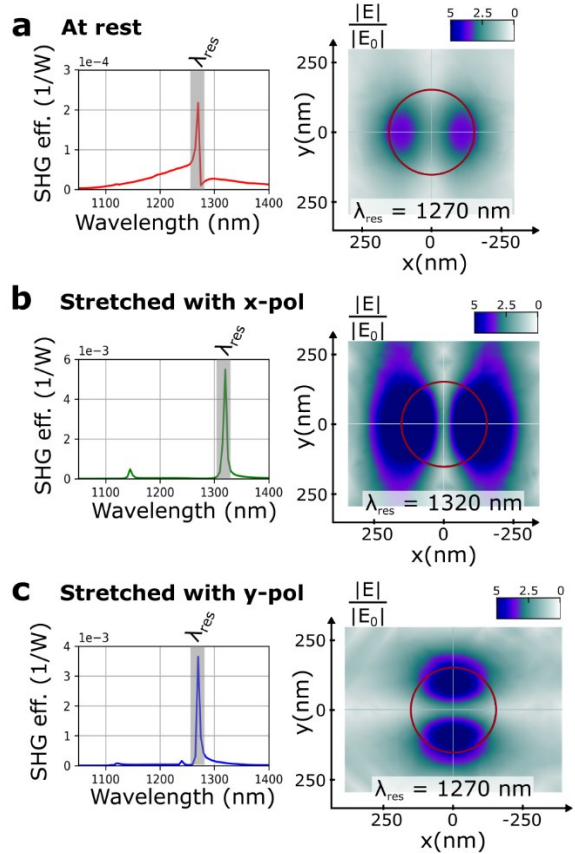

Figure S3. (a) SHG conversion efficiency spectrum for an array of NWs with length L = 900 nm, diameter D = 300 nm and pitch p = 600 nm. (b) At the resonant position λ = 1270nm, we plot the electric field distribution normalized to the incoming field. It is shown in xy-plane with the NW contour delimited by the red circle. (c-d) Similar plots for maximum resonance with x-polarization, which is for a pitch p = 690 nm and (e-f) for y-polarization which is for a pitch p = 780 nm. Field distribution as similar as in Figure 2a in the main text.

We computed the SHG conversion efficiency of an array of NWs with length L = 900 nm, diameter D = 300 nm and pitch p = 600 nm for which we changed only one parameter in order to observe the impact on the resonance. The length L is varied from 600 nm to 1100 nm, the diameter D from 260 nm to 300 nm and the whole structure is rescaled from 85% to 115% (the length, diameter and pitch are simultaneously changed by this factor) (see Figure S4). An increase in length or diameter will red-shift the resonant wavelength position and tune the intensity of the resonance (see Figure S4a,b), while rescaling the system will shift the resonance but preserve the maximum resonance intensity (see Figure S4c). It is possible to choose these three parameters to tune the resonant intensity and wavelength position.

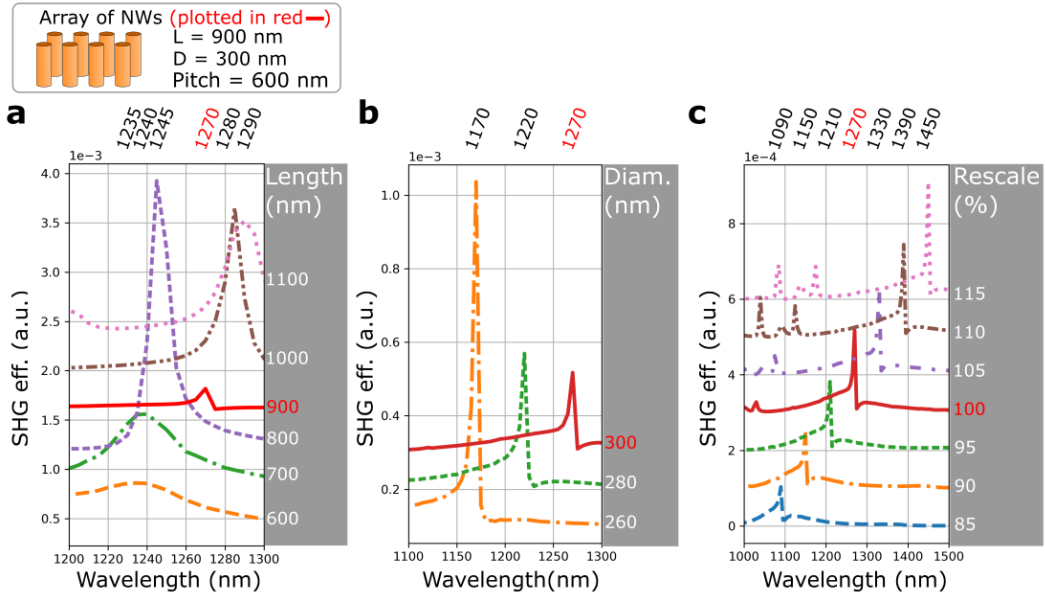

Figure S4. Resonance shift and intensity variation in an array of NWs for which we tuned (a) the length from 600 nm to 1100 nm, (b) the diameter from 260 nm to 300 nm and (c) rescaled the system from 85% to 115% (the length, diameter and pitch simultaneously by the same factor). An increase in length or diameter changes the resonance intensity and red-shifts its position, while rescaling the system only shifts the resonant position without affecting the intensity of the resonance.

## S2. Details on field distributions and resonances of single NWs

We compute with the finite element method simulations the scattering intensities and the multipole decomposition for a single nanowire (NW) with two different geometries of length L = 500 nm and diameter D = 260 nm (Figure S5) and of length L=900 nm and diameter 300 nm (Figure S6). We observe that for the smaller NW around a wavelength of 1100 nm the electric dipole (ED) and the magnetic dipole (MD) are dominant, while for the second NW, the ED and the MD dominate around 1400 nm. The spatial field distribution intensities of the electric $|E|^2$ and magnetic $|H|^2$ fields are shown in Figures S5b,c and S6b,c for both NWs at the wavelength corresponding to the dipole dominance. We can confirm this dominance by looking at the field nodes in the structure as we observe in figures S5,6 two nodes in the electric field distribution (at the top and bottom of the NW) and three nodes in the magnetic field distribution (top, bottom and around the middle of the NW).

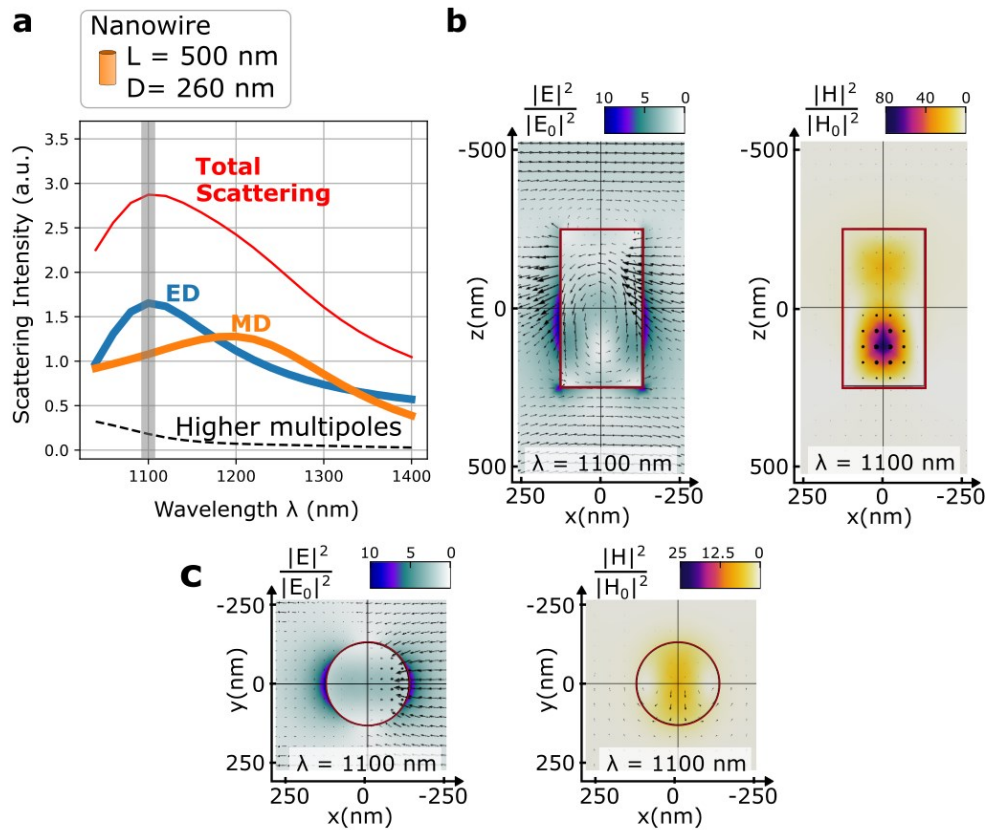

Figure S5. (a) Linear scattering spectrum of a NW with length 500 nm and diameter 260 nm. The NW is excited with a plane wave propagating in z-direction and polarized linearly in x-direction. The ED and MD dipoles are dominant around the wavelength 1100 nm. (b) Spatial field distribution intensities of the electric $|E|^2$ and magnetic $|H|^2$ fields shown in the xz-plane. The color scale gives the relative field enhancement and the black arrows are projections of the vectorial field. (c) Similar top view of spatial field distribution intensities. The shape of the NW is given with the red rectangle/circle. At 1100 nm, two nodes are present in the electric field distribution and three nodes in the magnetic field distribution, which indicates the dipole dominances.

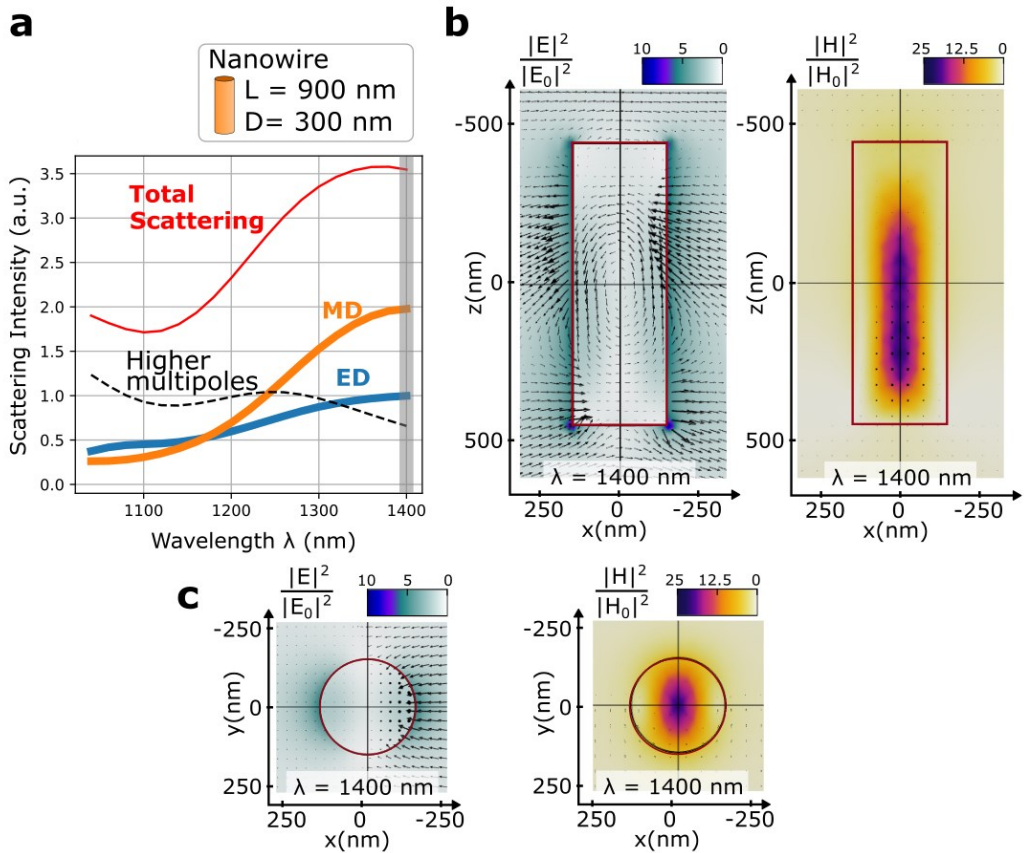

Figure S6. (a) Linear scattering spectrum from a single nanoantenna NW with length 900 nm and diameter 300 nm. The NW is excited with a plane wave propagating in z-direction and polarized linearly in x-direction. The ED and MD are dominant around the wavelength 1400 nm. (b) Spatial field distribution intensities of the electric $|E|^2$ and magnetic $|H|^2$ fields shown in the xz-plane. The color scale gives the relative field enhancement and the black arrows are projection of the vectorial field. (c) Similar top view of spatial field distribution intensities. The shape of the NW is given with the red rectangle/circle. At 1400 nm, two nodes are present in the electric field distribution and three nodes in the magnetic field distribution, which indicates the dipole dominances.

## S3 Detail on sample with NWs distribution of size

The fabrication procedure is described in the manuscript (see also Figure 4a). We show here the sample after metalorganic vapour-phase epitaxy (MOVPE) (see Figure S7a), after the Polydimethylsiloxane (PDMS) deposition (see Figure S7b) and after mechanical extraction with a Razor blade (see Figure S7c). We also capture two images to show the extracted NWs. The first Image, shown in Figure S7d, is obtained with a 5x objective (under light illumination), the second image, shown in Figure S7e, is obtained with a 100x objective (the contrast was expressed with a yellow/purple colormap). We can clearly observe with the 100x objective the periodicity of the structure given by the NWs. More details about the fabrication procedure are given in Methods.

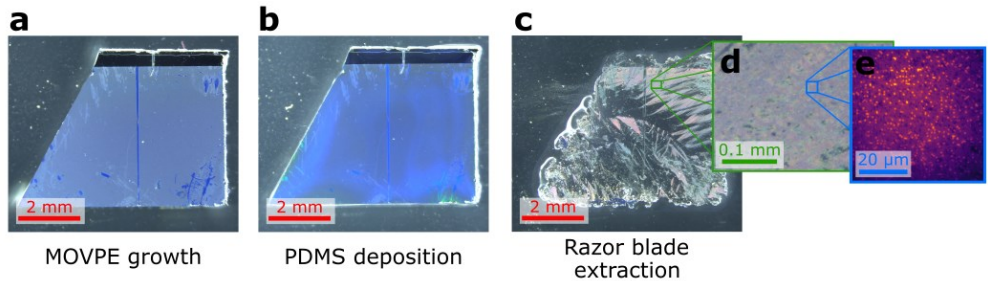

Figure S7. (a-c) Sample fabrication steps as given in Figure 4a from main text. Zoom in a region with extracted NWs obtained (d) with a 5x objective (under light illumination) and (e) with a 100x objective. The periodicity of the structure given by the NWs is clearly visible with the 100x objective.

The NWs composing the flexible photonic structure have varying sizes that we can describe with a distribution. For that we analysed several SEM images taken after MOVPE growth of the NWs and measured the lengths, diameters of the NWs and the pitch as the distance between NW centres. We observe that all NWs have the same diameter distribution but that NWs close to the edge are longer than in the middle of the sample. We can use this to characterize two regions where NWs have different lengths but similar diameters. The distributions are shown in Figure S8a-b for the two areas, with respectively a SEM image in the inset. The counts are described with a Weibull function as it fits best the asymmetric Gaussian-like distributions. The mean length for the shorter NWs is 560 ± 100 nm, while for the longer NWs it is 930 ± 125 nm. We can also observe that around 10% of NWs are not fully grown. The results for the radius are shown in Figure S8c and do not show a significant difference for different spots, the mean radius is 133 ± 18 nm.

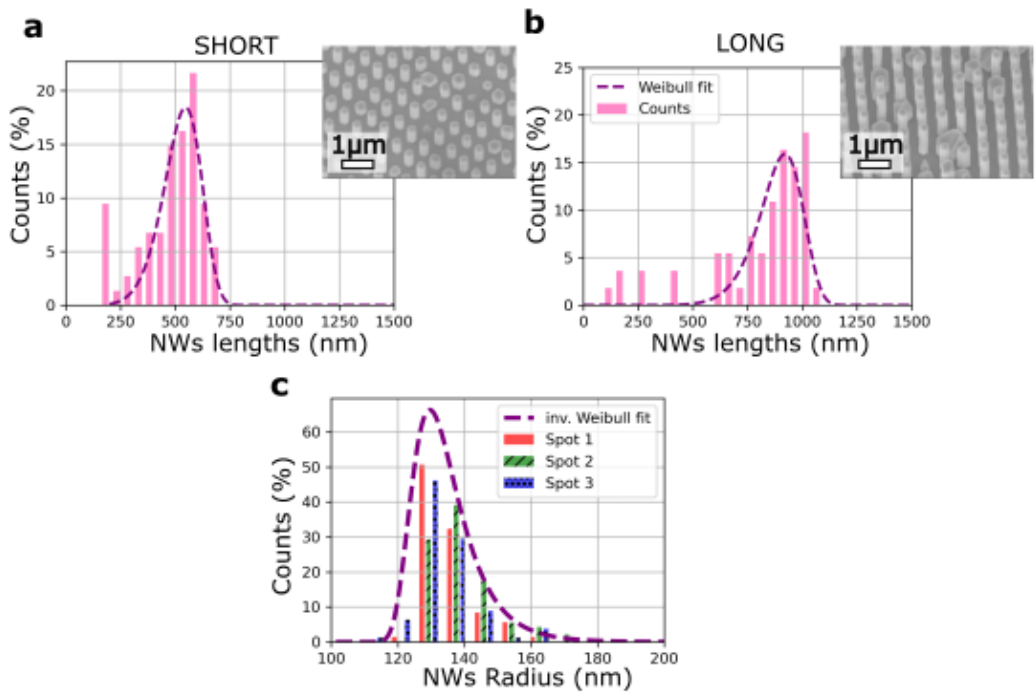

Figure S8. Distribution of length for the region with a) short NWs (around 500 nm length) and b) long NWs (around 900 nm length). Insets show the corresponding scanning electron microscopy images used to calculate the distribution of sizes. A Weibull function was taken to fit the distribution. c) Distribution of NWs radius shown for three different spots. The distribution is fitted with an inverted Weibull function. The distribution of diameter is constant over the whole sample.

The model used to compute into a single spectrum the SHG efficiency of an array with NWs of different sizes is explained in the Method section and illustrated here. Due to the not completely optimized bottom up growth method for the NWs, their heights and diameters varied locally as shown in Figure S8. A direct comparison between experimental results and simulations was not adequate as the latter was done for an ideal sample with NWs of unique height and diameter. As a first approximation, we calculated the discrete probability of having a NW with a certain size from corresponding SEM images, and added together the simulated spectrums of the ideal cases with their respective discrete weights. However, for the nonlinear SHG conversion efficiency spectrums, the IR resonance was too narrow and it would have needed a lot of simulations to avoid an unrealistic saw-like spectrum. Therefore we developped another method to come up with a single spectrum. Each SHG efficiency spectrum possesses a main peak that is fitted with a Gaussian function. We extracted for arrays with NWs of different lengths and diameters the amplitude (Figure S9a), the position (Figure S9b) and the width (Figure S9c) of this peak. The comparison between the reconstructed spectrum (in color, continuous line) and the original spectrum (in grey, dotted line) is shown for different arrays in Figure S9d. We came up with the single spectrum as Figure 5c,d in the main text by summing up the extrapolated spectrum with the corresponding probability distribution of the array with NWs the given length and diameter (see Figure S9e). This can be expressed like this: $\sum_{\text{Length, Diam}} \textit{Distribution} \cdot \textit{Amplitude} \cdot e^{-\left(\frac{\lambda - \textit{Position}}{\textit{Width}}\right)^2}$.

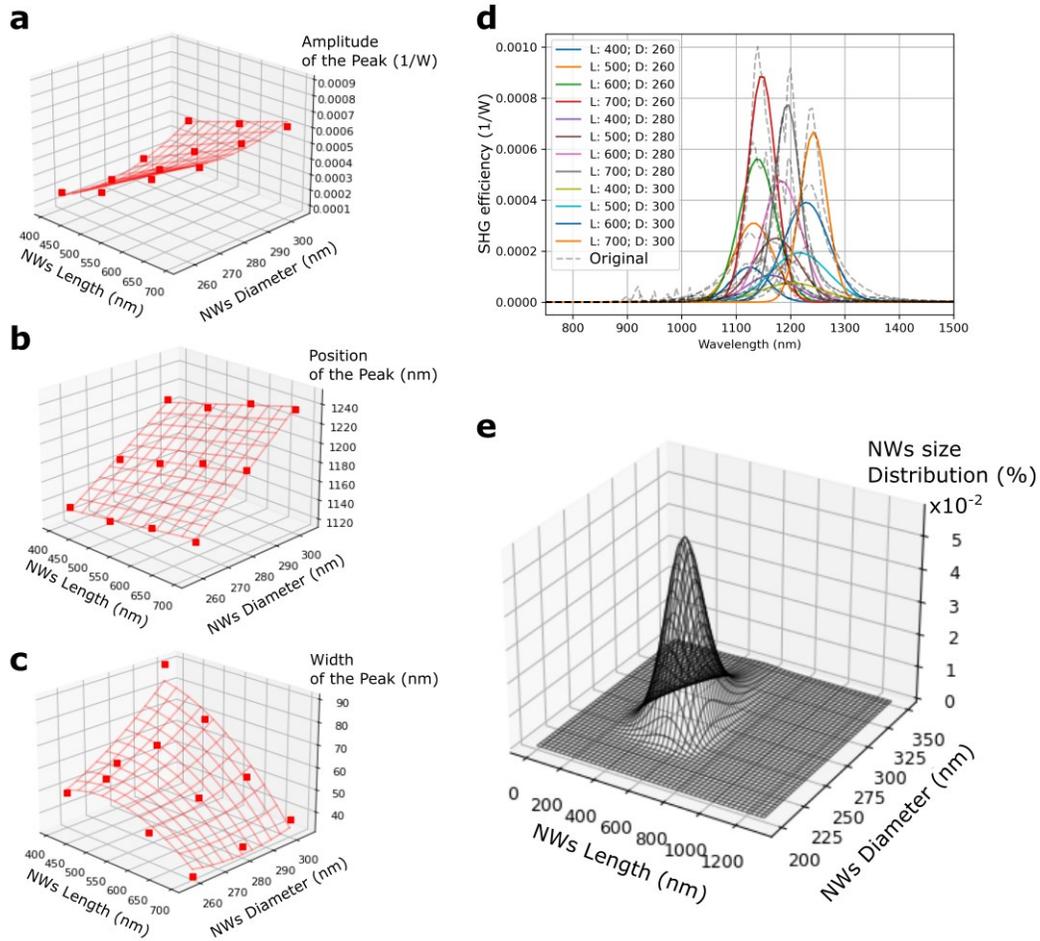

Figure S9. The SHG efficiency spectrum were fitted with a Gaussian function and (a) the amplitude, (b) position and (c) width of the main peak was recorded and extrapolated for every length and diameter of the NWs. (d) Comparison between the reconstructed spectrum (in color, continuous line) and the original spectrum (in grey, dotted line). (e) The probability of finding a NW with specific length and diameter for a specific region that was imaged with SEM.

## S4. Transmission measurement

We measured the linear transmission spectrum for the two different regions with short (around 500 nm length) and long NWs (around 900 nm length). The setup used here is the same one as used for this other linear measurement.[1] To simulate the transmission curve for the sample possessing a non-negligible distribution of sizes (NWs length and diameter), we summed up with the corresponding weight the transmission curves for an array of NWs with specific length and diameters. The experimental results and the simulation calculations are shown in Figure S10. We observe good matching in both cases as the shorter NWs have a large deep in transmission around 1100 nm while the longer NWs have a deep around 900 nm. The lower transmission observed experimentally can be explained by absorption and scattering in the PDMS or by the interaction of NWs with different sizes, which is not taken into account by the simulations model.

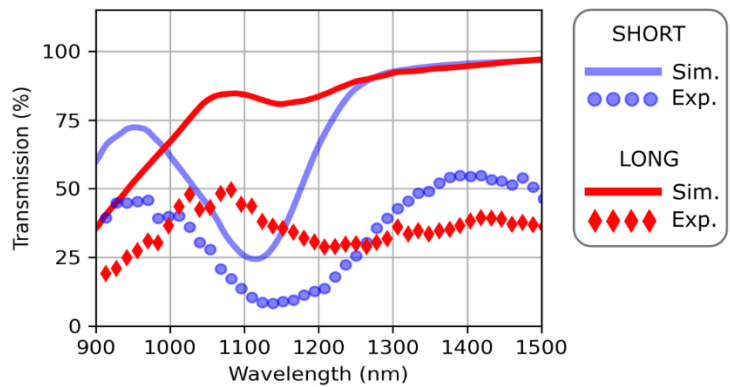

Figure S10. Experimental and simulation results of the transmission for two different regions on the sample with shorter (circle-blue) and longer NWs (diamonds-red). The experimental and simulation results agree well as the shorter NWs possess a transmission deep around 1100 nm while the longer NWs possess a transmission deep around 900nm.

## S5. Characterization of SHG signal

We provide an extensive characterization of the signal measured experimentally. We can observe a speckle pattern with the camera for every spot and both excitation polarization, see Figure S11a for x- and y-polarization. The Count rate recorded by the camera is above $10^3$ 1/s per pixel. The power dependence of the signal is also quadratic as expected for SHG signals (see Figure S11b). Figure S11c shows the spectrum of the excitation laser (in red), this excitation spectrum artificially frequency doubled (dotted) and the collected SHG signal (in blue). We clearly do not observe fluorescence from the sample.

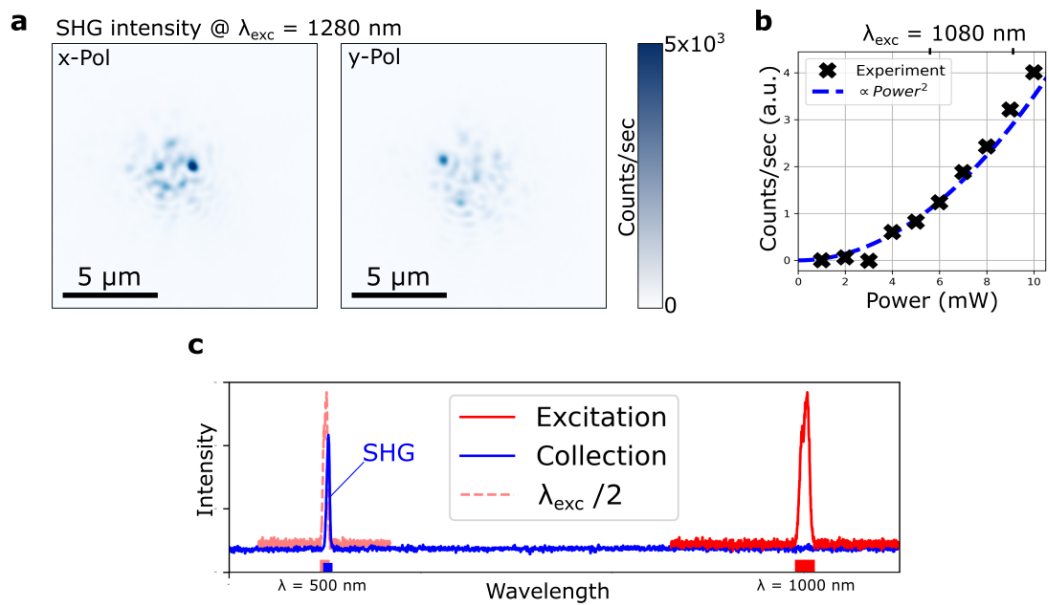

Figure S11. (a) Filtered image of the signal for the two polarizations: x-polarization and y-polarization. The camera counts (without any corrections) are on the order of $10^3$ 1/s. (b) Power dependence of the signal compared to a quadratic curve and (c) intensity spectrum of the excitation laser compared to the filtered signal. We clearly identify the signal as SHG.